\documentclass[prd,twocolumn,preprintnumbers]{revtex4-1}

\usepackage{amsmath}
\usepackage{epsfig}
\usepackage{graphicx}
\usepackage{color}
\usepackage[normalem]{ulem}
\usepackage[breaklinks, plainpages=false, colorlinks=true, anchorcolor=cyan, linkcolor=red, citecolor=cyan, urlcolor=magenta, bookmarks=false]{hyperref}

\usepackage[caption=false]{subfig}

\begin{document}
\renewcommand{\thefigure}{\arabic{figure}}
\setcounter{figure}{0}

 \def\I{{\rm i}}
 \def\E{{\rm e}}
 \def\D{{\rm d}}

\bibliographystyle{apsrev}

\title{Constraining the Polarization Content of Gravitational Waves with Astrometry}

\author{Logan O'Beirne}
\affiliation{eXtreme Gravity Institute, Department of Physics, Montana State University, Bozeman, Montana 59717, USA}

\author{Neil J. Cornish}
\affiliation{eXtreme Gravity Institute, Department of Physics, Montana State University, Bozeman, Montana 59717, USA}

\begin{abstract} 
Gravitational waves perturb the paths of photons, impacting both the time-of-flight and the arrival direction of light from stars. Pulsar timing arrays can detect gravitational waves by measuring the variations in the time of flight of radio pulses, while astrometry missions such as Gaia can detect gravitational waves from the time-varying changes in the apparent position of a field of stars. Just as gravitational waves impart a characteristic correlation pattern in the arrival times of pulses from pulsars at different sky locations,  the deflection of starlight is similarly correlated across the sky. Here we compute the astrometric correlation patterns for the full range of polarization states found in alternative theories of gravity, and decompose the sky-averaged correlation patterns into vector spherical harmonics. We find that the tensor and vector polarization states produce equal power in the electric- and magnetic-type vector spherical harmonics, while the scalar modes produce only electric-type correlations. Any difference in the measured electric and magnetic-type correlations would represent a clear violation of Einstein gravity. The angular correlations functions for the vector and scalar longitudinal modes show the same enhanced response at small angular separations that is familiar from pulsar timing.
\end{abstract}

\maketitle

\section{Introduction}

Gravitational wave astronomy has made it possible to test gravity in the dynamical, strong field regime. The first binary black hole~\cite{Abbott:2016blz} and binary neutron star~\cite{TheLIGOScientific:2017qsa} detections have been used to carry out a wide range of tests, including placing stringent bounds on the difference in propagation speed of gravity and light~\cite{Cornish:2017jml, Monitor:2017mdv}, constraining departures in the waveforms from the predictions of general relativity~\cite{TheLIGOScientific:2016src}, and constraining the polarization content of the signals~\cite{Abbott:2017tlp,Abbott:2017oio,Abbott:2018utx}. The existing array of ground based interferometers is not ideal for testing the polarization content of the signals as the two LIGO detectors and the single Virgo detector provide only a limited number of projections of the polarization. It takes at least 5 (mis-aligned) detectors to unambiguously resolve the polarization content~\cite{Chatziioannou:2012rf}. The pulsar timing approach to detecting gravitational waves is better suited to measuring polarization~\cite{2008ApJ685.1304L, daSilvaAlves:2011fp, Chamberlin:2011ev,Cornish:2017oic} as each pulsar line-of-sight provides a separate projection of the polarization state. The astrometric approach to detecting gravitational waves~\cite{Book:2010pf, Klioner:2017asb, PhysRevLett.119.261102} is similarly well suited to constraining the polarization content, with astrometric missions such as Gaia~\cite{2016A&A...595A...1G} observing billions of stars across the sky.

Photon trajectories are perturbed by gravitational waves, resulting in time delays and changes in the apparent position of the source. Gravitational waves cause the relative position of two stars on the sky to oscillate in a tensor correlation pattern. The formal expression for this tensor two-point correlation pattern, valid for any gravitational wave polarization, was first derived by Book and Flanagan~\cite{Book:2010pf}, however they only provided explicit expressions for the transverse-tensor modes of general relativity. Here we complete the derivation for the additional four polarization states that are possible in other metric theories of gravity~\cite{Will2014,lrr-2013-9}. Book and Flanagan~\cite{Book:2010pf} also derived the angular power spectrum, found by expanding the two-point correlation in terms of electric type (E) and magnetic type (B) vector spherical harmonics and integrating over the locations of the stars on the sky. This analysis is analogous to what is done when studying the polarization of the microwave background, where the polarization of the microwave photons are separated into curl-free E-modes and divergence-free B-modes. The tensor-transverse (TT) modes of general relativity produce identical angular correlation patterns for the EE and BB correlations at large angles, and differ slightly at small angular separations (the EB cross correlations vanish for all polarization states)~\cite{Book:2010pf}. Here we compute the angular power spectra for the additional non-Einsteinian polarization modes. The BB correlation vanishes for the scalar-transverse (ST) and the scalar-longitudinal (SL) polarizations while the EE correlation does not, indicating the the difference in the correlation functions $\Delta=({\rm EE}-{\rm BB})$ provides a powerful null test of general relativity. The vector-longitudinal (VL) polarizations evade this test since they produce identical EE and BB correlation patterns. Both the vector and scalar longitudinal modes produce an enhanced response at small angular separations that depends on the product of the gravitational wave frequency $f$ and the distance to the stars $L$. The enhancement is logarithmic for VL and linear for SL. Precisely the same enhancements occur in the two-point correlation functions for these modes for pulsar timing~\cite{2008ApJ685.1304L,  daSilvaAlves:2011fp, Chamberlin:2011ev}. This implies that astrometry missions like Gaia will be able to place much stronger constraints on the energy density in the longitudinal modes than for the transverse modes, as is the case for pulsar timing~\cite{Cornish:2017oic}.

In the final stages of writing this paper a preprint was posted with an independent calculation of the astrometric two-point correlation functions for non-Einsteinian polarization modes~\cite{Mihaylov:2018uqm}. After accounting for some differences in notation, we verified that our expressions for the two-point functions agree. Their treatment stopped short of computing the E and B mode angular power spectra, and those results are reported here for the first time. We work in geometrical units with $G=c=1$.

\section{Summary of Results}

Our calculation begins with the expression for the astrometric deflection $\delta {\bf n}({\bf n}, {\bf p}, t)$ for a star in the ${\bf n}$ direction perturbed by a gravitational wave source in the ${\bf p}$ direction. Book and Flanagan~\cite{Book:2010pf} provided a formal expression for the deflection that is valid for any gravitational wave polarization.  The next step is to compute the two-point correlation function $C^{ij}({\bf n}, {\bf n}', f)$  which describe the tensor deflection pattern produced by an isotropic stochastic background of gravitational waves.  Book and Flanagan~\cite{Book:2010pf} computed a formal expression for $C^{ij}({\bf n}, {\bf n}', f)$ that is valid for all polarizations. In practice, to make a detection, we need to combine measurements of the angular deflections between many pairs of stars since the angular deflections between any two stars is expected to be small relative to the noise in the measurement. This is analogous to measurements of the cosmic microwave background, where correlations are computed over large areas of the sky and expressed in terms of angular power spectra. To compute the angular power spectra of the astrometric deflections we decompose the angular deflections into electric-type (E) and magnetic-type (B) vector spherical harmonics and compute the correlations ${\rm EE, EB, BB}$ averaged over the distribution of stars in the sky. Assuming a uniform distribution of stars, the correlations only depend on the angle $\Theta ={\rm acos}({\bf n}\cdot {\bf n}')$ between the stars, or equivalently, the multipole moment $\ell$. 

\begin{figure}[htp]
\includegraphics[clip=true,angle=0,width=0.5\textwidth]{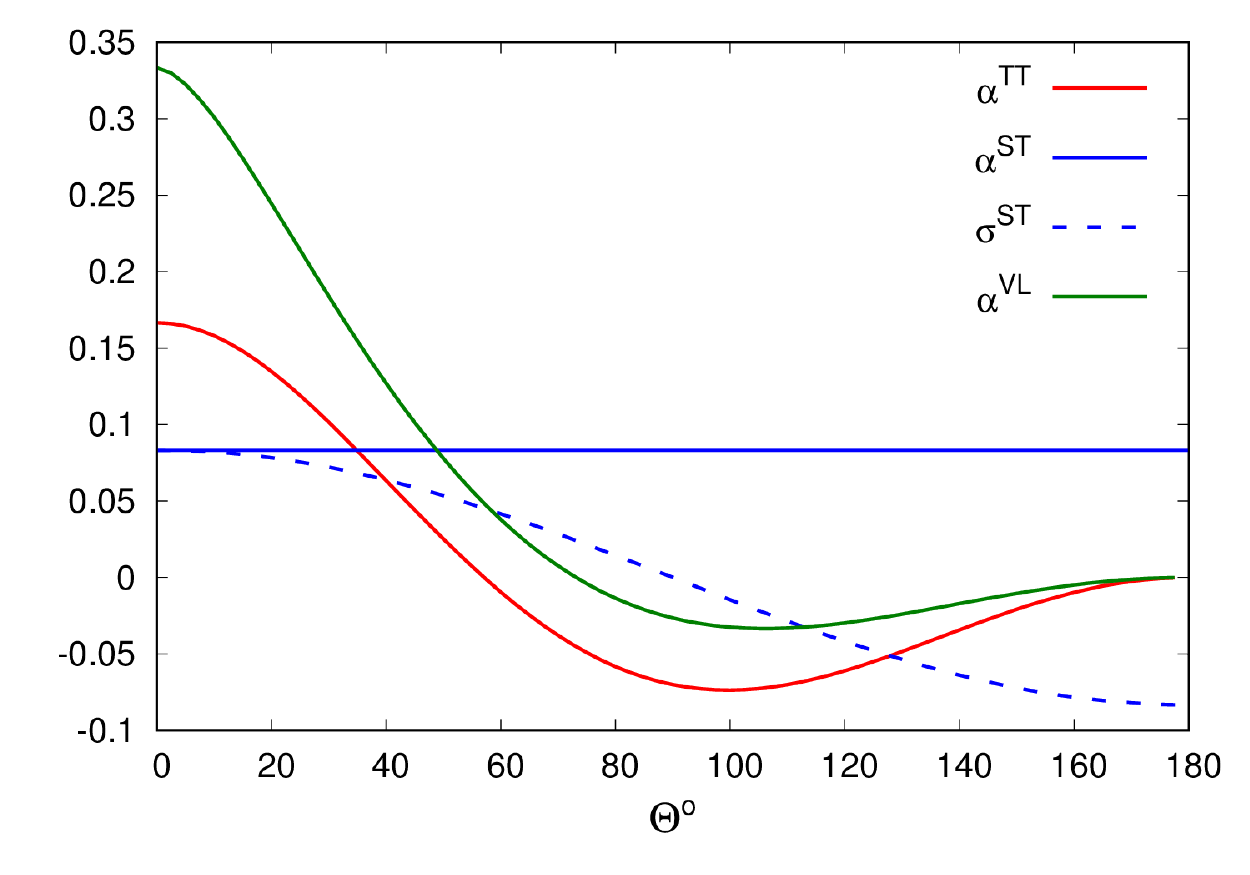} 
 \vspace*{-0.2in}
\caption{\label{fig:TTSTVL} The analytic correlation functions $\alpha(\Theta)$ and $\sigma(\Theta)$ for the TT, ST and VL modes.}
\end{figure}

The astrometric deflection $\delta {\bf n}({\bf n}, {\bf p}, t)$ depends on the phase of the gravitational wave at the Earth and at the star.  In the distant source limit, where the product  of the gravitational wave frequency $f$ and the distance to the star $L$ is large, the phase of the star term is uncorrelated between stars located at $L {\bf n}$ and $L' {\bf n}'$ unless $\Theta \approx 0$. For the transverse polarization states the star term can be safely ignored, but not so for the longitudinal modes. The general expression for the two-point correlation function for an isotropic stochastic background can be written as
\begin{eqnarray}
&& C^{ij}({\bf n}, {\bf n}', f) = \int d \Omega_{\bf p} \langle \delta n^i({\bf n},{\bf p},f) \delta n^i({\bf n}',{\bf p},f)  \rangle \nonumber \\
&& \quad = \frac{3 H_0^2}{32 \pi^2} \frac{\Omega_{\rm gw}(f)}{f^3} H^{ij}(\mathbf{n},\mathbf{n}')\, ,
\end{eqnarray}
where $H_0$ is the Hubble constant, $\Omega_{\rm gw}(f)$ is the energy density in gravitational waves per logarithmic frequency interval, scaled by the critical density $\rho_c = 3 H_0^2/(8 \pi)$,  $H^{ij}$ are the components of the deflection tensor
\begin{equation}\label{hij}
{\bf H}(\mathbf{n},\mathbf{n}') = \alpha(\Theta) ({\bf a} \otimes {\bf a})- \sigma(\Theta) ({\bf b} \otimes {\bf c}),
\end{equation}
and the unit triad ${\bf a}, {\bf b}, {\bf c}$ is defined by ${\bf a} = ({\bf n} \times {\bf n} ')/\sin\Theta$,  ${\bf b} = ({\bf n} ({\bf n}\cdot {\bf n} ') - {\bf n} ')/\sin\Theta$, ${\bf c} = ({\bf n} '({\bf n}\cdot {\bf n} ') - {\bf n} )/\sin\Theta$.  The angular correlations functions for the tensor transverse (TT), scalar transverse (ST) and vector longitudinal (VL) modes can be computed in closed form in the distant star limit $fL \rightarrow \infty$, and are given by
\begin{eqnarray}\label{large}
&& \alpha^{\rm TT} =  \sigma^{\rm TT } = (1/12)(7\,\text{cos}\,\Theta-5)\nonumber \\
&&  \hspace*{0.8in} -8\,\text{ln}(\text{sin}(\Theta/2))(\text{sin}^{6}(\Theta/2)/\text{sin}^{2}\Theta)\nonumber \\
&& \alpha^{\rm ST} =  \frac{1}{12}  \nonumber \\
&& \sigma^{\rm ST} =  \frac{1}{12}\text{cos}(\Theta)  \nonumber \\
&& \alpha^{\rm VL} =  \sigma^{\rm VL } =  \frac{\text{sin}^{2}(\Theta/2)}{\text{sin}^{2}(\Theta)}\Big[8 \, \text{sin}^{2}(\Theta/2) \text{ln}(\text{sin}(\Theta/2) \nonumber \\
&& \hspace*{0.9in} + \frac{1}{3}\Big(3+2\,\text{cos}(\Theta)-\text{cos}(2\Theta)\Big)\Big]\, .
\end{eqnarray}
Plots of the TT, ST and VL correlation patterns are shown in Figure~\ref{fig:TTSTVL}.

The scalar longitudinal (SL) case is more complicated as the star terms need to be included to arrive at a finite expression at small angular separations. We were unable to derive a closed-form expression for the SL correlation functions that is valid for all angles, and instead quote results that are valid for $\Theta=0$ and $\Theta \gg 1/(fL)$. For large angles we have
\begin{eqnarray}\label{SLlarge}
&&  \alpha^{\rm SL} \approx - \frac{ \text{sin}^{2}(\Theta/2)}{3\,\text{sin}^{2}(\Theta)} \Big[ 1+\text{cos}(\Theta)+3\,\text{ln}(\text{sin}(\Theta/2)) \Big] \nonumber \\
&& \sigma^{\rm SL} \approx  - \frac{ \text{sin}^{2}(\Theta/2)}{6\,\text{sin}^{2}(\Theta)} \Big[ 4+5\,\text{cos}(\Theta) +\text{cos}(2\Theta) \nonumber \\
&& \hspace*{1.4in}  +6\,\text{ln}(\text{sin}(\Theta/2)) \Big] \, .
\end{eqnarray}
The divergence at $\Theta=0$ in these expressions is regularized to a logarithmic dependence on the distance to the stars when the star terms are included. To leading order in $fL \gg 1$ we find
\begin{eqnarray}\label{zero}
&&  \alpha^{\rm SL}(0) = \sigma^{\rm SL}(0) = \frac{1}{8} \Big[\frac{ \Phi_{s}^2 - {\Phi'_{s}}^2}{4 \Phi_{s} \Phi'_{s}}\ln(\Phi_{s}^2- {\Phi'_{s}}^2) \nonumber \\
&& \hspace*{0.4in}  +   \ln(\Phi_{s} \Phi'_{s}) -\frac{ \Phi_{s}^2 + {\Phi'_{s}}^2}{4 \Phi_{s} \Phi'_{s}}\ln(\Phi_{s} +\Phi'_{s}) +\dots \Big],
\end{eqnarray}
where $\Phi_{s} = 2\pi fL$ and $\Phi'_{s} = 2\pi fL'$. 
 \begin{figure}[htp]
\includegraphics[clip=true,angle=0,width=0.5\textwidth]{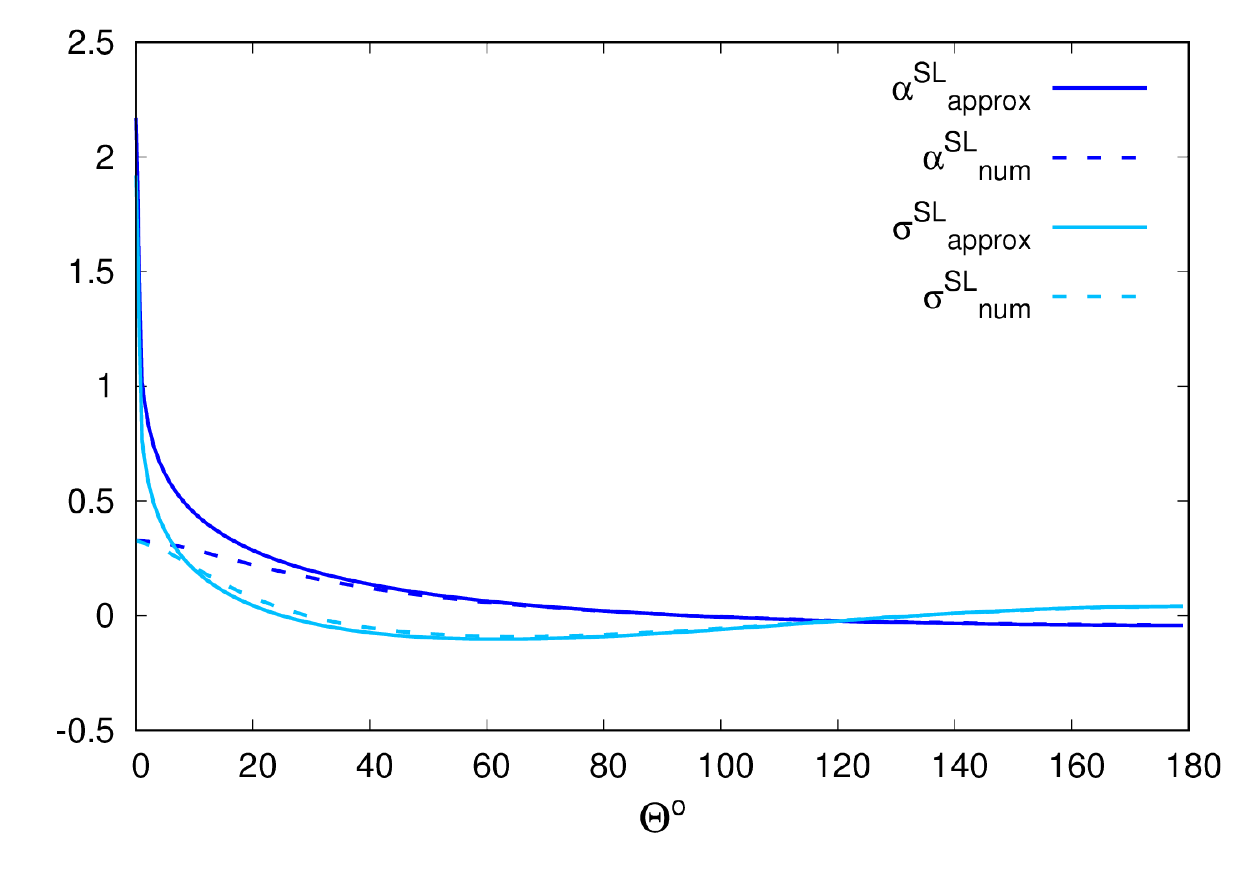} 
 \vspace*{-0.2in}
\caption{\label{fig:SL} The correlation functions $\alpha(\Theta)$ and $\sigma(\Theta)$ for the SL mode. The solid lines show the analytic approximation, valid for $\Theta \gg 1/(fL)$. The dotted lines are found by numerical integration for the case $fL = fL'=10$.}
\end{figure}

The angular power spectra can be expressed in terms of the multipole coefficients
\begin{equation}
C^{QQ'}_{\ell m \ell' m'} = \int d^2\Omega_{\mathbf{n}} d^2\Omega_{\mathbf{n}'} Y^{Q*}_{\ell mi}(\mathbf{n}) Y^{Q'}_{\ell 'm'j}(\mathbf{n}')  H_{ij}(\mathbf{n},\mathbf{n}').
\end{equation}
Here $Q,Q'$ refer to the electric-type and magnetic-type vector spherical harmonics
\begin{eqnarray}\label{vsph}
 &&\mathbf{Y}^E_{lm}(\mathbf{n}) = (\ell(\ell+1))^{-1/2} \mathbf{\nabla} Y_{\ell m}(\mathbf{n}) \nonumber \\
&&  \mathbf{Y}^B_{lm}(\mathbf{n}) = (\ell(\ell+1))^{-1/2} (\mathbf{n}\times\mathbf{\nabla}) Y_{\ell m}(\mathbf{n})
 \end{eqnarray}
The EB cross-correlations vanish due to the different parity of the E and B modes. For an isotropic distribution of stars we find
\begin{equation}
C^{QQ'}_{\ell m \ell' m'} =  \delta^{QQ'}\delta_{\ell \ell'} \delta_{m m'} \frac{1}{\ell (\ell+1)} \frac{4 \pi}{2\ell+1} {\rm QQ}_\ell \, ,
\end{equation}
where
\begin{equation}
 {\rm QQ}_\ell = \int d\Theta \, P_\ell(\cos\Theta)  {\rm QQ}(\Theta) \, .
\end{equation}
The formal expressions for the EE and BB correlation functions are:
\begin{eqnarray}\label{EEBB}
 && {\rm EE}(\Theta) =  \nabla_i \nabla'_j \left[ H^{ij}(\mathbf{n},\mathbf{n}') \right] \nonumber \\
&&  {\rm BB}(\Theta) = \nabla^l {\nabla'}^p \left[\epsilon_{ikl}\epsilon_{jmp} n^k {n'}^m H^{ij}(\mathbf{n},\mathbf{n}') \right].
 \end{eqnarray} 
 Note that ${\rm EE}(\Theta)$ and ${\rm BB}(\Theta)$, or equivalently ${\rm EE}_\ell$ and  ${\rm BB}_\ell$, are physical observables that can be inferred from astrometric data, just as the EE and BB power spectra for the CMB polarization are physical observables. A method for extracting $C^{QQ'}_{\ell m \ell' m'}$, and hence ${\rm EE}_\ell$ and ${\rm BB}_\ell$, from Gaia observations is described in Ref.~\cite{Mignard:2012xm}.

 \begin{figure}[htp]
\includegraphics[clip=true,angle=0,width=0.5\textwidth]{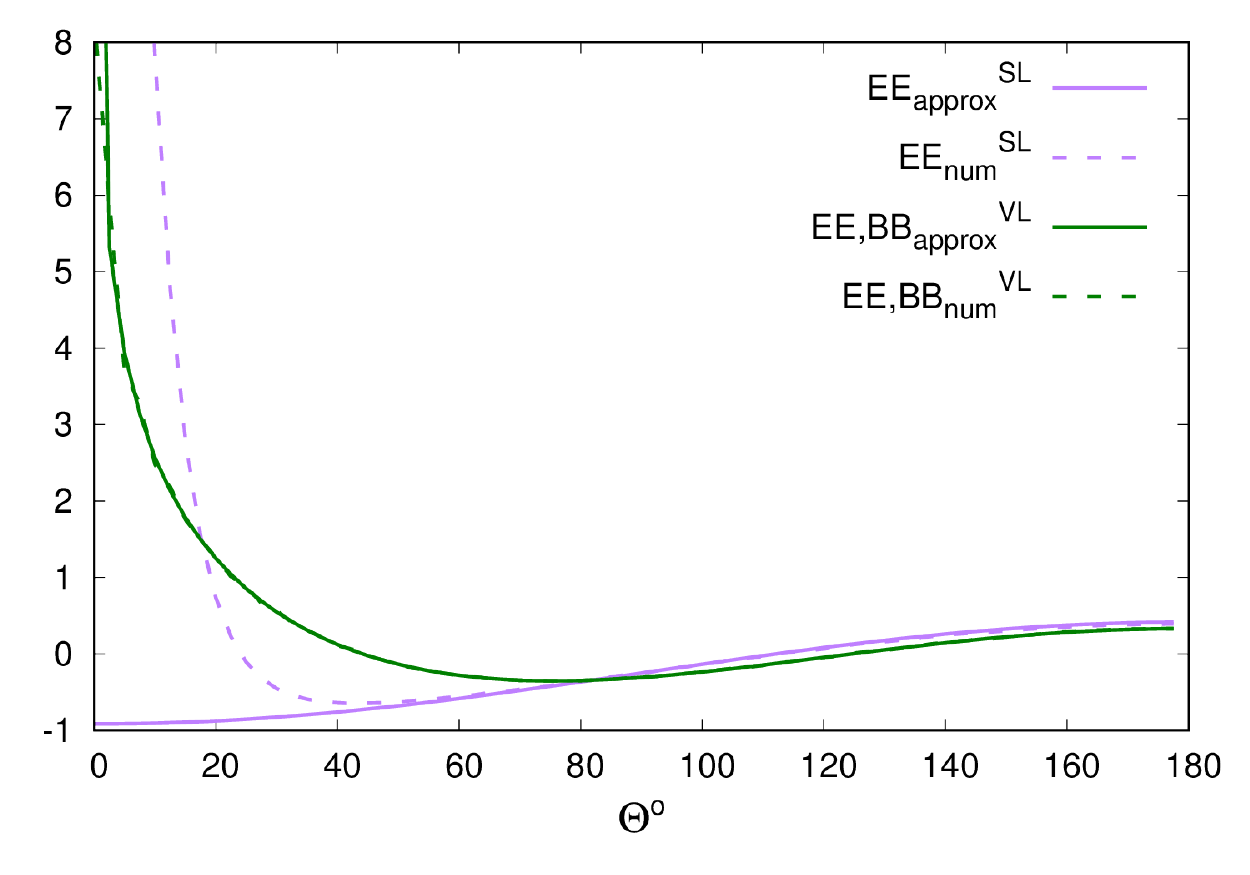} 
 \vspace*{-0.2in}
\caption{\label{fig:EEBBL} The upper plot shows EE and BB angular correlation functions for the longitudinal modes. The solid lines are the analytic approximations from Eq. (\ref{EB}) and the dotted lines are from a numerical evaluation with $fL=10$.}
\end{figure}

At large angles we find that ${\rm EE}(\Theta) = {\rm BB}(\Theta)$ for the TT and VL modes, and for the scalar modes ST and SL we find ${\rm BB}(\Theta) = 0$, as expected from the even parity of the scalar perturbations. Additional care must be taken when computing the angular power spectra as the derivatives acting on $H_{ij}(\mathbf{n},\mathbf{n}')$ act on the phase of the star term, and modify the behavior at small angles. In the distant star limit we find
 \begin{eqnarray}\label{EB}
&& {\rm EE}^{\rm TT}(\Theta) = \frac{2}{3} \Big[2 - \Big(\frac{1-\text{cos}\Theta}{2} \Big) \nonumber \\
&& \hspace*{0.8in} + 6\,\Big(\frac{1-\text{cos}\Theta}{2} \Big)\text{ln}\Big(\frac{1-\text{cos}\Theta}{2} \Big) \Big] \nonumber \\
&& {\rm BB}^{\rm TT}(\Theta) =  {\rm EE}^{\rm TT}(\Theta) \nonumber  \\
&& {\rm EE}^{\rm VL}(\Theta)  \approx -\frac{1}{3} \Big(3+ 4\,\text{cos}(\Theta) + 6\, \text{ln}(\text{sin}(\Theta/2))\Big) \nonumber \\
&& {\rm BB}^{\rm VL}(\Theta) =  {\rm EE}^{\rm VL}(\Theta) \nonumber \\
&& {\rm BB}^{\rm ST}(\Theta) = {\rm BB}^{\rm SL}(\Theta) =  0  \nonumber \\
&& {\rm EE}^{\rm ST}(\Theta) = \frac{1}{3}\,\text{cos}\Theta \nonumber \\
&& {\rm EE}^{\rm SL}(\Theta) \approx -\frac{ 1}{12}(3+ 8\,\text{cos}\,\Theta),
\end{eqnarray}
Note that up to an overall scaling, the correlation pattern for the TT mode is {\it identical} to the Hellings-Downs curve from pulsar timing. Similarly, the EE correlation for the ST mode has the same form as the ST correlation function from pulsar timing, aside from an overall constant offset and scaling.
The expressions for the scalar longitudinal and vector longitudinal modes are only valid for $\Theta \gg 1/(fL)$.  Derivatives of the star terms become important at small angular separations. A complete numerical treatment would need to integrate over the radial distribution of stars, but to simplify the calculation we will assume that all the stars are at the same distance.  Figure~\ref{fig:EEBBL} compares the closed-form approximations to the correlation functions for the longitudinal modes to the numerical evaluation with $fL=10$. 

\begin{figure}[htp]
\includegraphics[clip=true,angle=0,width=0.5\textwidth]{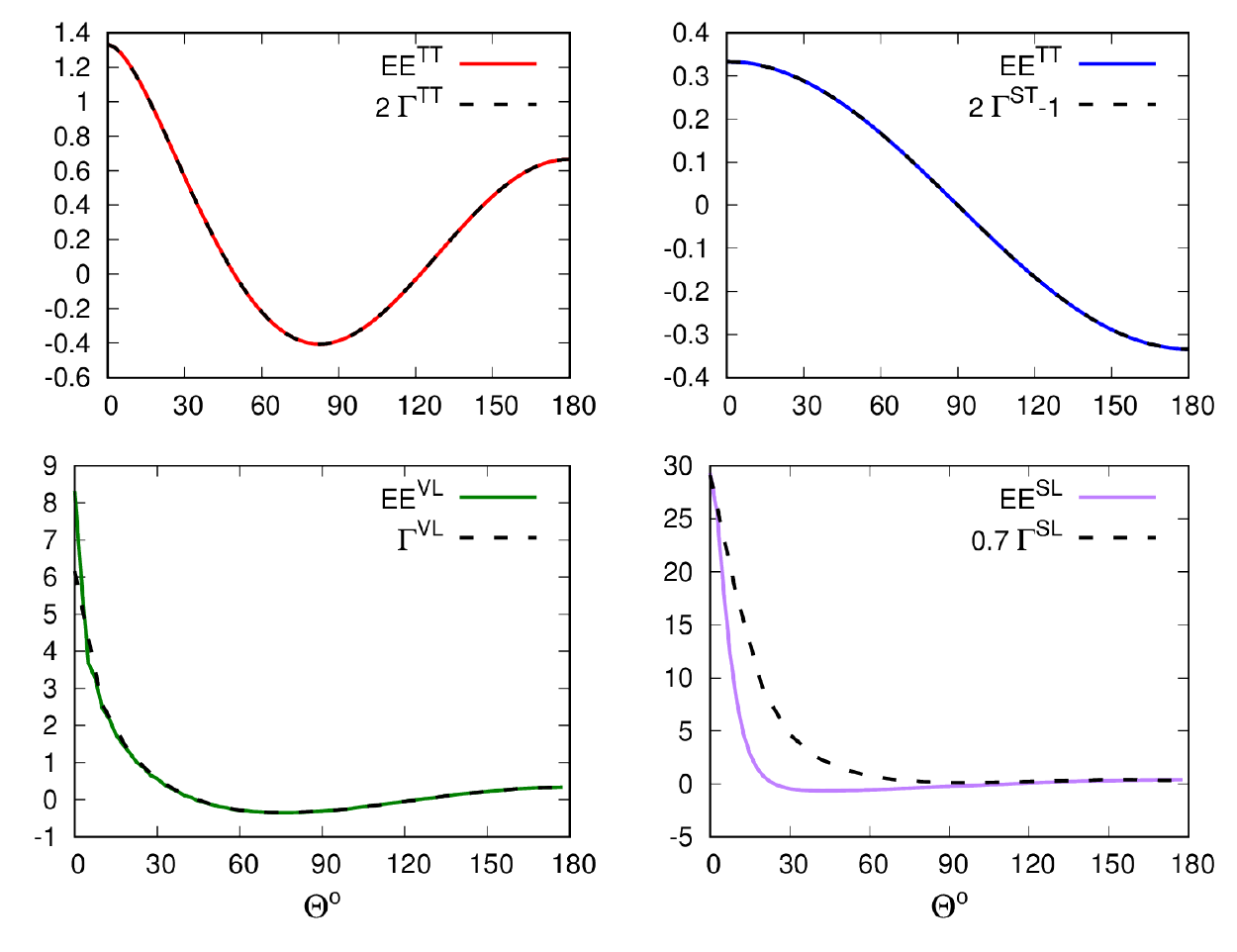} 
 \vspace*{-0.2in}
\caption{\label{fig:EEBBT} The EE angular correlation functions are compared to the corresponding correlation functions $\Gamma$ for pulsar timing. Aside from overall scalings and offsets, the correlation functions for the transverse modes are identical. The vector longitudinal correlation functions are identical at large angles, but differ slightly at small angles, though both share a similar $\ln(fL)$ scaling at $\Theta=0$ . The correlation functions for the scalar longitudinal modes are qualitatively similar, but differ in their detailed form. They share a similar $fL$ dependent scaling at $\Theta=0$. The plots are shown for $fL=10$.}
\end{figure}

The derivatives of $H_{ij}(\mathbf{n},\mathbf{n}')$ that appear in the definition of the EE and BB correlation functions enhance the importance of the star terms at small angular separations and produce an enhanced response. For the VL modes we find that the enhancement at $\Theta=0$ scales as $\ln(fL)$, while for the SL mode we find that the enhancement scales as $fL$. These enhancement factors are identical to those found for the VL and SL angular correlation functions for pulsar timing. This implies that astrometric observatories such as Gaia will be able to place much tighter constraints on the energy density of the longitudinal modes, as is the case for pulsar timing~\cite{Cornish:2017oic}. Figure~\ref{fig:EEBBT} compares the numerically computed EE correlation functions to the corresponding two-point angular correlation functions $\Gamma$ for pulsar timing~~\cite{2008ApJ685.1304L}. Aside from an overall difference in scaling and a constant shift for the scalar mode, the correlation functions for the transverse modes are {\em identical} for astrometry and pulsar timing. The vector longitudinal correlation functions are identical at large angles, but differ slightly at small angles, though both scale as $\ln(fL)$ at small angles, albeit with different scaling coefficients. The correlation functions for the scalar longitudinal modes are qualitatively similar, but differ in their detailed form. They both scale as $fL$ for small angles, but with different scaling coefficients.

\section{Computing the two-point correlation functions}

The angular deflection of a star at location ${\bf N} = {\bf n} L$ caused by a plane gravitational wave $h_{ij}(t,\textbf{x}) = \Re[\mathcal{H}_{ij}e^{-2 \pi i f (t - {\bf p}\cdot{\bf x})} ]$ propagating in the ${\bf p}$ direction is given by~\cite{Book:2010pf}
\begin{equation}
 \delta n^i({\bf n}, {\bf p}, t) =\Re\Big[ \mathcal{R}^{ikl}(\textbf{n},\textbf{p}) \mathcal{H}_{kl}\text{e}^{-2 \pi i f t }\Big]
\end{equation}
where 
\begin{equation}
\mathcal{R}^{ikl}(\textbf{n},\textbf{p}) = \frac{1}{2} \Big[ \frac{(T_{\textbf{n}_{1}}\,n^{i} + T_{\textbf{p}}\,p^{i})\,n^{k}\,n^{l}}{1 + \textbf{p}\cdot\textbf{n}} - T_{\textbf{n}_{2}}\,n^{k} \delta^{il} \Big]
\end{equation}
and
\begin{eqnarray}
&& T_{\textbf{n}_{1}} = 1-\frac{i(2+{\bf p}\cdot{\bf n}) }{\Phi_s (1+{\bf p}\cdot{\bf n})} e^{i \Phi_s (1+{\bf p}\cdot{\bf n})}  \nonumber \\
&& T_{\textbf{p}} = 1- \frac{i}{\Phi_s (1+{\bf p}\cdot{\bf n})} e^{i \Phi_s (1+{\bf p}\cdot{\bf n})}) \nonumber \\
&& T_{\textbf{n}_{2}} =  \frac{1}{2} - \frac{i}{\Phi_s (1+{\bf p}\cdot{\bf n})} e^{i \Phi_s (1+{\bf p}\cdot{\bf n})}
\end{eqnarray}
with $\Phi_s = 2 \pi f L$. To compute the two-point correlation function $C^{ij}({\bf n}, {\bf n}', f)$  we define the orthonormal coordinate system $({\bf m}, {\bf l}, {\bf n})$ and introduce the gravitational wave polarization tetrad $({\bf u}, {\bf v}, {\bf p})$:
\begin{eqnarray}
&& {\bf p} = \cos\theta\, {\bf n} + \sin\theta(\cos\phi\, {\bf m} + \sin\phi\, {\bf l}),  \nonumber \\
&& {\bf u} = \frac{\left(-\sin\theta\cos\phi\, {\bf n} + \cos\theta\, {\bf m}\right)}{\sqrt{\sin^2\theta \cos^2\phi +\cos^2\theta}},\nonumber \\
&& {\bf v} =- \frac{\left( \sin\theta \cos\theta \sin\phi\, {\bf n}   + \sin^2\theta\cos\phi\sin\phi\, {\bf m} \right)}{\sqrt{\sin^2\theta \cos^2\phi +\cos^2\theta}} \nonumber \\
&& \hspace*{0.4in}  + \frac{ (\sin^2\theta \cos^2\phi+\cos^2\theta)\, {\bf l}}{\sqrt{\sin^2\theta \cos^2\phi +\cos^2\theta}} 
\end{eqnarray}
The basis tensors for the various gravitational wave polarization states are then
\begin{eqnarray}
&&\boldsymbol{\epsilon}^+_{\rm TT} = {\bf u} \otimes {\bf u}  - {\bf v} \otimes {\bf v}\nonumber \\
&&\boldsymbol{\epsilon}^\times_{\rm TT} = {\bf u} \otimes {\bf v}  + {\bf v} \otimes {\bf u}\nonumber \\
&&\boldsymbol{\epsilon}^\odot_{\rm ST} = {\bf u} \otimes {\bf u}  + {\bf v} \otimes {\bf v}\nonumber \\
&&\boldsymbol{\epsilon}^u_{{\rm VL}} = {\bf u} \otimes {\bf p}  + {\bf p} \otimes {\bf u}\nonumber \\
&&\boldsymbol{\epsilon}^v_{{\rm VL}} = {\bf v} \otimes {\bf p}  + {\bf p} \otimes {\bf v}\nonumber \\
&&\boldsymbol{\epsilon}^\odot_{{\rm SL}} = {\bf p} \otimes {\bf p} \, .
\end{eqnarray}
The ${\bf n}'$  vector can be written as 
\begin{equation}
{\bf n}' = \cos\Theta\, {\bf n} + \sin\Theta(\cos\Phi\, {\bf m} + \sin\Phi\, {\bf l}) \, .
\end{equation}
The two-point correlation function $C^{ij}({\bf n}, {\bf n}', f)$ can be written in terms of the tensor $H^{ij}(\mathbf{n},\mathbf{n}') = \alpha(\Theta) a^i a^j - \sigma(\Theta) b^i c^j $. Written in the  the $({\bf m}, {\bf l}, {\bf n})$ coordinate system we have
\begin{eqnarray}
&&{\bf a} =  ( -\sin\Phi\, {\bf m} +  \cos\Phi\, {\bf l}) \nonumber \\
&& {\bf b} = - ( \cos\Phi\, {\bf m} +  \sin\Phi\, {\bf l})  \nonumber \\
&&{\bf c} = -\sin\Theta\, {\bf n} + \cos\Theta(\sin\Phi\, {\bf m} + \cos\Phi\, {\bf l}) \, .
\end{eqnarray}
The angular correlation functions are given by
\begin{eqnarray}
&&\alpha^{P}(\Theta) = \frac{1}{4\pi\,\text{sin}^{2}(\Theta)} \int d^{2}\Omega_{\bf p}a^{i}\mathcal{R}_{ikl}(\textbf{n},\textbf{p})\,\epsilon^{P}_{kl}(\textbf{p}) \nonumber \\
&& \hspace*{1.2in} \times \, a^{j}\mathcal{R}_{jrs}(\textbf{n}',\textbf{p})^{*}\,\epsilon^{P}_{rs}(\textbf{p}) \nonumber \\
&&\sigma^{P}(\Theta) = -\frac{1}{4 \pi \,\text{sin}^{2}(\Theta)} \int d^{2}\Omega_{\bf p}b^{i}\mathcal{R}_{ikl}(\textbf{n},\textbf{p})\,\epsilon^{P}_{kl}(\textbf{p}) \nonumber \\
&& \hspace*{1.2in} \times \,  c^{j}\mathcal{R}_{jrs}(\textbf{n}',\textbf{p})^{*}\,\epsilon^{P}_{rs}(\textbf{p})
\end{eqnarray}
where $d^2\Omega_{\mathbf{p}} = \sin\theta \, d\theta d\phi$ .
The correlation functions for the TT and VL modes are defined as being summed over the $+,\times$ or $u,v$ states. The correlation functions for ST and SL only have a single contribution. The full expression for the correlation functions are lengthy. To render the expressions manageable we follow Ref.~\cite{Book:2010pf} and introduce the quantities $\kappa = \mathbf{n}\cdot\mathbf{p}$, $\kappa' = \mathbf{n}'\cdot\mathbf{p}$, $\lambda = \mathbf{n}\cdot\mathbf{n}'$, $\nu^{2} = 1 - \kappa^{2} $, $\nu'^{2} = 1 - \kappa'^{2} $, $\mu = \sin\Theta\, \mathbf{a}\cdot\mathbf{p}$, which satisfy the identity $1 + 2\lambda\kappa\kappa' = \mu^{2} + \lambda^{2} + \kappa^{2} + \kappa'^{2} $. The correlation functions are then given by
\begin{eqnarray}
	&&\alpha^{\rm TT}(\Theta) = \frac{1}{16\pi \sin^2(\Theta)} \int d^2\Omega_{\mathbf{p}} \Big[ \frac{\mu^{2}(\nu^{2}\nu'^{2}-2\mu^{2})}{(1+\kappa)(1+\kappa')} T_{\textbf{p}}T_{\textbf{p}'}^{*}  \nonumber \\
	&&   \hspace*{1.1in}-\frac{\mu^{2}(\kappa'\kappa^{2}-2\lambda\kappa+\kappa')}{(1+\kappa)}T_{\textbf{p}}T_{\textbf{n}_{2}'}^{*}\nonumber\\
	&&  \hspace*{1.1in}-\frac{\mu^{2}(\kappa\kappa'^{2}-2\lambda\kappa'+\kappa)}{(1+\kappa')}T_{\textbf{n}_{2}}T_{\textbf{p}'}^{*}\nonumber\\
	&& \hspace*{1.0in}   +(\lambda-\kappa\kappa')(1-\lambda^{2}-\mu^{2})T_{\textbf{n}_{2}}T_{\textbf{n}_{2}'}^{*} \Big]
\end{eqnarray}

\begin{eqnarray}
&&\sigma^{\rm TT}(\Theta) = -\frac{1}{16\pi \sin^2(\Theta)} \int d^2\Omega_{\mathbf{p}}\times\nonumber \\
&& \hspace*{0.3in}\Big[ \frac{(\lambda\kappa-\kappa')(\lambda\kappa'-\kappa)(\nu^{2}\nu'^{2}-2\mu^{2})}{(1+\kappa)(1+\kappa')} T_{\textbf{p}}T_{\textbf{p}'}^{*}  \nonumber \\
&&   -\frac{(\lambda\kappa-\kappa')[\lambda(\nu^{2}\nu'^{2}-2\mu^{2})-\nu^{2}(\lambda-\kappa\kappa')]}{(1+\kappa)}T_{\textbf{p}}T_{\textbf{n}_{2}'}^{*}\nonumber\\
&&   -\frac{(\lambda\kappa'-\kappa)[\lambda(\nu^{2}\nu'^{2}-2\mu^{2})-\nu'^{2}(\lambda-\kappa\kappa')]}{(1+\kappa')}T_{\textbf{n}_{2}}T_{\textbf{p}'}^{*}\nonumber\\
&& \hspace*{0.6in} +[\lambda^{2}(\nu^{2}\nu'^{2}-2\mu^{2}-\nu^{2}-\nu'^{2})\nonumber\\
&&  \hspace*{0.8in} +\lambda\kappa\kappa'(\nu^{2}+\nu'^{2})+\nu^{2}\nu'^{2}]T_{\textbf{n}_{2}}T_{\textbf{n}_{2}'}^{*} \Big]
\end{eqnarray}

\begin{eqnarray}
&&\alpha^{\rm ST}(\Theta) = \frac{1}{16\pi \sin^2(\Theta)} \int d^2\Omega_{\mathbf{p}} \Big[ \mu^{2}(1-\kappa)(1-\kappa') T_{\textbf{p}}T_{\textbf{p}'}^{*}  \nonumber \\
&&  \hspace*{0.7in} +\mu^{2}\kappa'(1-\kappa)T_{\textbf{p}}T_{\textbf{n}_{2}'}^{*}+\mu^{2}\kappa(1-\kappa')T_{\textbf{n}_{2}}T_{\textbf{p}'}^{*}\nonumber \\
&& \hspace*{1.8in} + (\mu^{2} \kappa \kappa') T_{\textbf{n}_{2}}T_{\textbf{n}_{2}'}^{*} \Big]
\end{eqnarray}

\begin{eqnarray}
&&\sigma^{\rm ST}(\Theta) = -\frac{1}{16\pi \sin^2(\Theta)} \int d^2\Omega_{\mathbf{p}}\times\nonumber \\
&& \hspace*{0.2in}\Big[ (\lambda\kappa-\kappa')(\lambda\kappa'-\kappa)(1-\kappa)(1-\kappa') T_{\textbf{p}}T_{\textbf{p}'}^{*}  \nonumber \\
&&  \hspace*{0.5in} +(\lambda\kappa-\kappa')(\lambda\kappa'-\kappa)(1-\kappa)\kappa'T_{\textbf{p}}T_{\textbf{n}_{2}'}^{*}\nonumber\\
&&  \hspace*{0.5in} +(\lambda\kappa-\kappa')(\lambda\kappa'-\kappa)(1-\kappa')\kappa T_{\textbf{n}_{2}}T_{\textbf{p}'}^{*}\nonumber\\
&& \hspace*{0.8in} +[\lambda^{2}\nu^{2}\nu'^{2}-\lambda(\lambda-\kappa\kappa')(\nu^{2}+\nu'^{2})\nonumber\\
&&  \hspace*{1.6in} +(\lambda-\kappa\kappa')^{2}]T_{\textbf{n}_{2}}T_{\textbf{n}_{2}'}^{*} \Big]
\end{eqnarray}

\begin{eqnarray}
&&\alpha^{\rm VL}(\Theta) = \frac{1}{16\pi \sin^2(\Theta)} \int d^2\Omega_{\mathbf{p}} \frac{4\mu^{2}\kappa\kappa'(\lambda-\kappa\kappa')}{(1+\kappa)(1+\kappa')} T_{\textbf{p}}T_{\textbf{p}'}^{*} \nonumber\\
&&   -\frac{2\mu^{2}\kappa(\lambda-2\kappa\kappa')}{(1+\kappa)}T_{\textbf{p}}T_{\textbf{n}_{2}'}^{*}-\frac{2\mu^{2}\kappa'(\lambda-2\kappa\kappa')}{(1+\kappa')}T_{\textbf{n}_{2}}T_{\textbf{p}'}^{*}\nonumber\\
&& \hspace*{0.7in}   +[\lambda\mu^{2}-\kappa\kappa'(4\mu^{2} + \lambda^{2} - 1)]T_{\textbf{n}_{2}}T_{\textbf{n}_{2}'}^{*} \Big]
\end{eqnarray}

\begin{eqnarray}
&&\sigma^{\rm VL}(\Theta) = -\frac{1}{16\pi \sin^2(\Theta)} \int d^2\Omega_{\mathbf{p}}\times\nonumber \\
&& \hspace*{0.3in}\Big[ \frac{4(\lambda\kappa-\kappa')(\lambda\kappa'-\kappa)\kappa\kappa'(\lambda-\kappa\kappa')}{(1+\kappa)(1+\kappa')} T_{\textbf{p}}T_{\textbf{p}'}^{*}  \nonumber \\
&&   -\frac{2(\lambda\kappa-\kappa')[2\lambda\kappa\kappa'(\lambda-\kappa\kappa')-\kappa(\kappa'-\kappa'\kappa^{2}+\lambda\kappa)]}{(1+\kappa)}T_{\textbf{p}}T_{\textbf{n}_{2}'}^{*}\nonumber\\
&&   -\frac{2(\lambda\kappa'-\kappa)[2\lambda\kappa\kappa'(\lambda-\kappa\kappa')-\kappa'(\kappa-\kappa\kappa'^{2}+\lambda\kappa')]}{(1+\kappa')}T_{\textbf{n}_{2}}T_{\textbf{p}'}^{*}\nonumber\\
&&  \hspace*{0.2in} +[(\kappa^{2} - \kappa'^{2})^{2} + \lambda^{2}(\lambda^{2} - 1) + \mu^{2}(\mu^{2} - 1)]T_{\textbf{n}_{2}}T_{\textbf{n}_{2}'}^{*} \Big]
\end{eqnarray}

\begin{eqnarray}
&&\alpha^{\rm SL}(\Theta) = \frac{1}{16\pi \sin^2(\Theta)} \int d^2\Omega_{\mathbf{p}} \Big[ \frac{\mu^{2}\kappa^{2}\kappa'^{2}}{(1+\kappa)(1+\kappa')} T_{\textbf{p}}T_{\textbf{p}'}^{*}  \nonumber \\
&&  \hspace*{1.0in} -\frac{\mu^{2}\kappa'\kappa^{2}}{(1+\kappa)}T_{\textbf{p}}T_{\textbf{n}_{2}'}^{*}-\frac{\mu^{2}\kappa\kappa'^{2}}{(1+\kappa')}T_{\textbf{n}_{2}}T_{\textbf{p}'}^{*}\nonumber \\
&& \hspace*{1.9in} + \mu^{2} \kappa \kappa' T_{\textbf{n}_{2}}T_{\textbf{n}_{2}'}^{*} \Big]
\end{eqnarray}

\begin{eqnarray}
&&\sigma^{\rm SL}(\Theta) = -\frac{1}{16\pi \sin^2(\Theta)} \int d^2\Omega_{\mathbf{p}}\times\nonumber \\
&& \hspace*{0.8in}\Big[ \frac{(\lambda\kappa-\kappa')(\lambda\kappa'-\kappa)\kappa^{2}\kappa'^{2}}{(1+\kappa)(1+\kappa')} T_{\textbf{p}}T_{\textbf{p}'}^{*}  \nonumber \\
&&  \hspace*{0.8in} -\frac{(\lambda\kappa-\kappa')(\lambda\kappa'-\kappa)\kappa^{2}\kappa'}{(1+\kappa)}T_{\textbf{p}}T_{\textbf{n}_{2}'}^{*}\nonumber\\
&&  \hspace*{0.8in} -\frac{(\lambda\kappa'-\kappa)(\lambda\kappa-\kappa')\kappa'^{2}\kappa}{(1+\kappa')}T_{\textbf{n}_{2}}T_{\textbf{p}'}^{*}\nonumber\\
&&  \hspace*{0.3in} +[(\lambda^{2} + 1)\kappa^{2}\kappa'^{2} - \lambda\kappa\kappa'(\kappa^{2}+\kappa'^{2})]T_{\textbf{n}_{2}}T_{\textbf{n}_{2}'}^{*} \Big]
\end{eqnarray}
with
\begin{eqnarray}
&& T_{\textbf{n}_{1}} = 1-\frac{i(2+\kappa) }{\Phi_s (1+\kappa)} e^{i \Phi_s (1+\kappa)}  \nonumber \\
&& T_{\textbf{p}} = 1- \frac{i}{\Phi_s (1+\kappa)} e^{i \Phi_s (1+\kappa)}) \nonumber \\
&& T_{\textbf{n}_{2}} =  \frac{1}{2} - \frac{i}{\Phi_s (1+\kappa)} e^{i \Phi_s (1+\kappa)} \nonumber \\
&& T_{\textbf{n}_{1}'} = 1-\frac{i(2+\kappa') }{\Phi'_s (1+\kappa')} e^{i \Phi_s (1+\kappa')}  \nonumber \\
&& T_{\textbf{p}'} = 1- \frac{i}{\Phi'_s (1+\kappa')} e^{i \Phi_s (1+\kappa')}) \nonumber \\
&& T_{\textbf{n}_{2}'} =  \frac{1}{2} - \frac{i}{\Phi'_s (1+\kappa')} e^{i \Phi'_s (1+\kappa')} \, .
\end{eqnarray}
In the distant star limit, $\Phi_s, \Phi'_s \gg 1$, and away from $\theta=\pi$, the phase terms oscillate rapidly and can be discarded, allowing the integrals to be evaluated in closed form to give the expressions quoted in Eqs.(\ref{large},\ref{SLlarge}). Without the phase terms, the $\alpha$ and $\sigma$ integrands for SL mode diverge at $\theta=\pi$. Restoring the phase terms renders the integrands finite, but the integrals are then intractable, save at $\Theta=0$, where the leading order terms scale logarithmically with $fL$, as shown in Eq.(\ref{zero}).

\section{Computing the angular power spectra}

\begin{figure}[htp]
\includegraphics[clip=true,angle=0,width=0.5\textwidth]{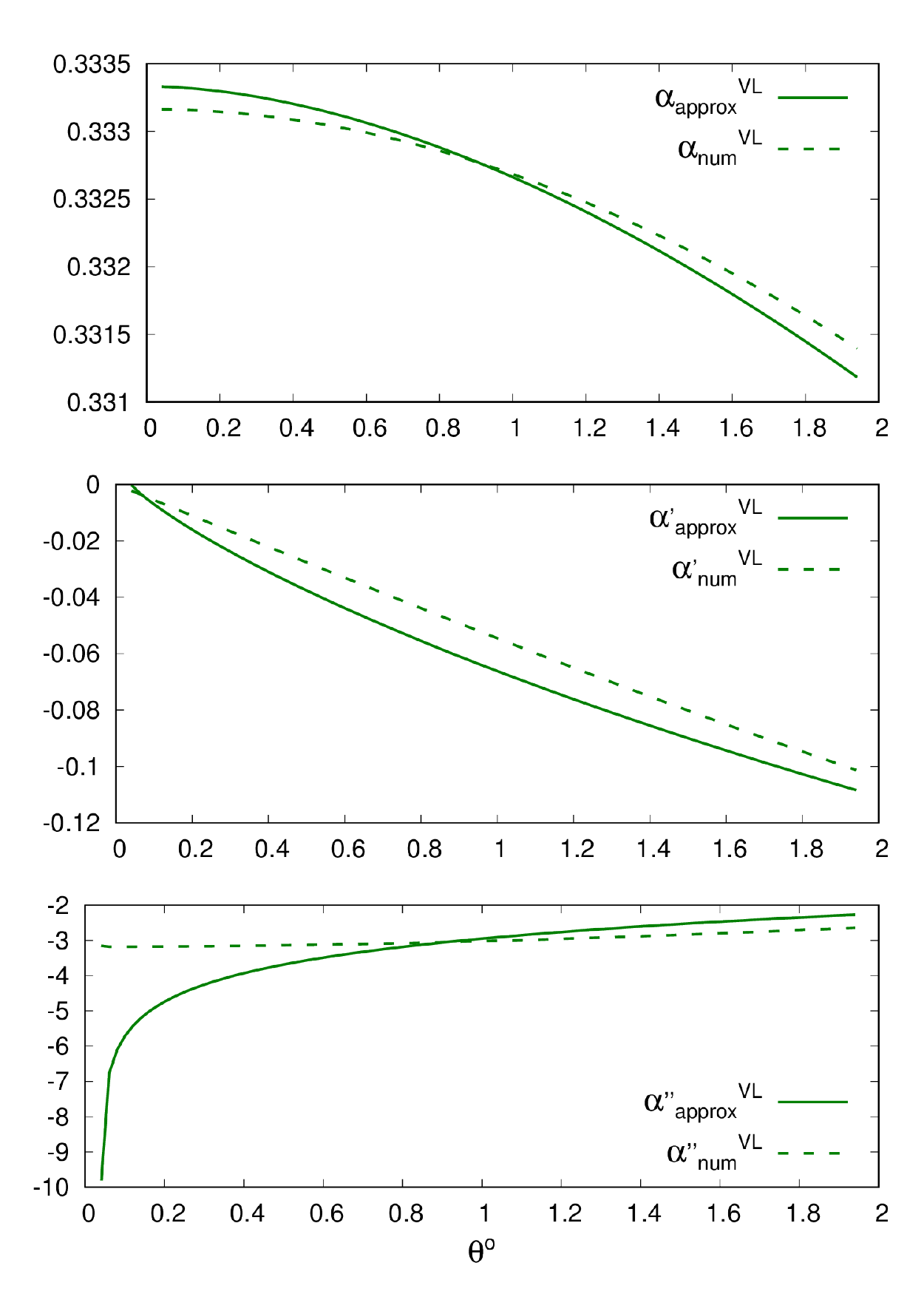} 
 \vspace*{-0.2in}
\caption{\label{fig:alpha} Comparing the analytic expressions without the star term for $\alpha^{\rm VL}(\Theta), \alpha'^{\rm VL}(\Theta), \alpha''^{\rm VL}(\Theta)$ to a numerical evaluation with the star terms illustrates the differences that occur at small angles, which h are especially significant for the $\alpha''^{\rm VL}(\Theta)$ term. The star terms play an important role in determining the behavior of the ${\rm EE}(\Theta)$ and ${\rm BB}(\Theta)$ correlation functions for small angles. The plots are shown for $fL=10$.}
\end{figure}

In contrast with pulsar timing, where the small number of millisecond pulsars allow searches for gravitational waves to be performed directly in terms of the two-point correlation functions, the large number of stars available for an astrometric search make it more natural to integrate over the distribution of stars on the sky and compute angular correlation functions, as is done with the cosmic microwave background. To achieve this, the astrometric deflections $\delta {\bf n}$ are first decomposed in terms of vector spherical harmonics:
\begin{equation}
 \delta n^i({\bf n}, {\bf p}, f) = \sum_{\ell m} \delta n_{E \ell m}(f) \mathbf{Y}^E_{\ell m}(\mathbf{n}) + \delta n_{B \ell m}(f) \mathbf{Y}^B_{\ell m}(\mathbf{n}).
\end{equation}
The angular power spectra are found by integrating the two-point correlation function $C^{ij}({\bf n}, {\bf n}', f)$ over a uniform distribution of stars ${\bf n}$, ${\bf n}'$:
\begin{eqnarray}
&& \Gamma^{QQ'}_{\ell m \ell' m'}(f) = \int d^2\Omega_{\mathbf{n}} d^2\Omega_{\mathbf{n}'} 
 Y^{Q*}_{\ell mi}(\mathbf{n}) Y^{Q'}_{\ell' m'j}(\mathbf{n}') C^{ij}({\bf n}, {\bf n}', f) \nonumber \\
 &&  \hspace*{0.65in} = \frac{3 H_0^2 \Omega_{\rm gw}(f)}{16 \pi^3 f^3}  \, \delta^{QQ'} \delta_{\ell \ell'} \delta_{m m'} C^Q_{\ell}(f) \, ,
\end{eqnarray}
Here $Q=E,B$ label the electric and magnetic terms, and $C^Q_\ell(f)$ defines the angular power spectrum
\begin{eqnarray}
 C^Q_\ell(f) &= &\int d^2\Omega_{\mathbf{n}} d^2\Omega_{\mathbf{n}'} 
 Y^{Q*}_{\ell mi}(\mathbf{n}) Y^{Q}_{\ell m j}(\mathbf{n}') H^{ij}(\mathbf{n},\mathbf{n}') \nonumber \\
 &=&  \frac{4\pi}{2\ell+1} \int d(\text{cos}\,\Theta) \, P_\ell(\cos\Theta)  {\rm QQ}(\Theta) \, .
\end{eqnarray}
To arrive at this expression, the derivatives of the ordinary spherical harmonics that appear in the definitions of the vector spherical harmonics are converted to derivatives acting on the angular
displacement tensor $H_{ij}$ by integrating by parts, resulting in the formal expressions:
\begin{eqnarray}\label{EEBB2}
 && {\rm EE}(\Theta) =  \nabla_i \nabla'_j \left[ H^{ij}(\mathbf{n},\mathbf{n}') \right] \nonumber \\
&&  {\rm BB}(\Theta) = \nabla^l {\nabla'}^p \left[\epsilon_{ikl}\epsilon_{jmp} n^k {n'}^m H^{ij}(\mathbf{n},\mathbf{n}') \right].
 \end{eqnarray} 
Recalling the expression for $H_{ij}$ from Eq. (\ref{hij}), we see that the derivatives will act on both the basis tensors and the $\alpha,\sigma$ correlation functions. Writing $\tilde{H}_{lp} = \epsilon_{ikl}\epsilon_{jmp} n_k n'_m H^{ij}$
we find that
\begin{equation}\label{hij}
\tilde {\bf H}(\mathbf{n},\mathbf{n}') = \sigma(\Theta) ({\bf a} \otimes {\bf a})- \alpha(\Theta) ({\bf b} \otimes {\bf c}).
\end{equation}
For the TT and VL modes $\alpha=\sigma$, so that $\tilde{H}_{ij} ={H}_{ij}$ and ${\rm EE}(\Theta)={\rm BB}(\Theta)$. For the ST and SL modes $\alpha\neq \sigma$ and ${\rm EE}(\Theta)\neq {\rm BB}(\Theta)$.  Indeed, a direct calculation shows that the magnetic-type correlation vanishes for the scalar modes, ${\rm BB}^{\rm ST}(\Theta) = {\rm BB}^{\rm SL}(\Theta) = 0$, which follows from the even parity of the scalar perturbations. Evaluating the derivatives that appear in Eq. (\ref{EEBB2}) we find
\begin{eqnarray}
	&& \nabla_{i}a^{i}a^{j} = \frac{(\textbf{n}\cdot\textbf{n}')n'^{j}-n^{j}}{\text{sin}^{2}(\Theta)} \nonumber \\
	&& \nabla_{i}b^{i}c^{j} = \frac{(\textbf{n}\cdot\textbf{n}')[(\textbf{n}\cdot\textbf{n}')n'^{j}-n^{j}]}{\text{sin}^{2}(\Theta)} \nonumber \\
	&& \nabla_{i}\nabla'_{j} a^{i}a^{j} =\frac{2(\textbf{n}\cdot\textbf{n}')}{\text{sin}^{2}(\Theta)} \nonumber \\
	&& \nabla_{i}\nabla'_{j} b^{i}c^{j} =  - 1 \nonumber \\
	&& \nabla_{i}\sigma(\Theta) = -\sigma'(\Theta)\frac{n'_{i} - (\textbf{n}\cdot\textbf{n}')n_{i}}{\text{sin}(\Theta)}\nonumber \\
	&& \nabla'_{j}\sigma(\Theta) = -\sigma'(\Theta)\frac{n_{j} - (\textbf{n}\cdot\textbf{n}')n'_{j}}{\text{sin}(\Theta)} \nonumber \\
	&& \nabla_{i}\nabla'_{j} \sigma(\Theta)= \sigma'(\Theta) \Bigg\{\frac{\delta_{ij}-n_{i}n_{j} - n'_{i}n'_{j} + (\textbf{n}\cdot\textbf{n}')n_{i}n'_{j}}{-\text{sin}(\Theta)} \nonumber \\
	&& \hspace*{0.4in} + \frac{\text{cos}(\Theta)[n'_{i}-(\textbf{n}\cdot\textbf{n}')n_{i}][n_{j}-(\textbf{n}\cdot\textbf{n}')n'_{j}]}{-\text{sin}^{3}(\Theta)}\Bigg\} \nonumber \\
	&& \hspace*{0.4in} +\sigma''(\Theta)\frac{[n'_{i}-(\textbf{n}\cdot\textbf{n}')n_{i}][n_{j}-(\textbf{n}\cdot\textbf{n}')n'_{j}]}{\text{sin}^{2}(\Theta)} 
\end{eqnarray} 
and similarly for $\alpha(\Theta)$. Here the primes denote derivatives with respect to $\Theta$. Combining these pieces together yields
\begin{equation}\label{EEBBfull}
\begin{split}
{\rm EE}(\Theta) = -\sigma''(\Theta) + \frac{1}{\text{sin}\,\Theta}\alpha'(\Theta) - 2\frac{\text{cos}\,\Theta}{\text{sin}\,\Theta}\sigma'(\Theta)+\sigma(\Theta)\\
{\rm BB}(\Theta) = -\alpha''(\Theta) + \frac{1}{\text{sin}\,\Theta}\sigma'(\Theta) - 2\frac{\text{cos}\,\Theta}{\text{sin}\,\Theta}\alpha'(\Theta)+\alpha(\Theta)
\end{split}
\end{equation}
Care needs to be exercised in evaluating these expressions: each of the terms are separately infinite at $\Theta=0$, and, moreover, the derivatives also act on the ``star'' terms, which changes the character of the integrand over the the source direction ${\bf p}$. The longitudinal terms can no longer be evaluated in closed form for all $\Theta$. Ignoring the star terms yield the expressions quoted in 
Eq.(\ref{EB}). These expressions are valid everywhere for the transverse modes, but are only valid for $\Theta \gg 1/(fL)$ for the longitudinal modes.  Figure~\ref{fig:alpha} compares analytic and numerical evaluations of $\alpha^{\rm VL}(\Theta), \alpha'^{\rm VL}(\Theta), \alpha''^{\rm VL}(\Theta)$. The numerical evaluation includes the contribution from the star terms, while the analytic expressions do not. We see that star term plays an important role in determining the small angle behavior of $\alpha''^{\rm VL}(\Theta)$. The same if true for $\alpha''(\Theta)$ and $\sigma''(\Theta)$ all the polarization modes.

The correlation functions for the longitudinal modes are enhanced at small angles.

To study the small angle behavior of the correlation functions in more detail we re-write Eq.(\ref{EEBBfull}) in terms of $\bar\alpha(\Theta, \bf p)$ and $\bar\sigma(\Theta, \bf p)$, where $\alpha(\Theta) = \int  d \Omega_{\bf p}\, \bar\alpha(\Theta, \bf p)$ and similarly for $\sigma(\Theta)$. The correlation functions are then given by ${\rm EE}(\Theta) = \int  d \Omega_{\bf p}\, {\rm ee}(\Theta, \bf p)$ and similarly for BB, where the integrands ${\rm ee}(\Theta, \bf p)$ and  ${\rm bb}(\Theta, \bf p)$ are finite at $\Theta=0$. Using these expressions we find several interesting results: First, the EE and BB correlation functions for the transverse traceless modes are {\em not} equal at small angles since
\begin{eqnarray}
&& {\rm EE}^{\rm TT}(\Phi_{s})|_{\Theta=0} =  \frac{22}{15} + \mathcal{O}\Big(\frac{1}{\Phi_{s}^{2}} \Big)  \nonumber\\
&& {\rm BB}^{\rm TT}(\Phi_{s})|_{\Theta=0} =  \frac{8}{3} + \mathcal{O}\Big(\frac{1}{\Phi_{s}^{2}} \Big) \, .
\end{eqnarray}
The fact that ${\rm EE}^{\rm TT}(\Theta)\neq {\rm BB}^{\rm TT}(\Theta)$ at small angles was missed by Book and Flanagan~\cite{Book:2010pf} since they neglected the star terms. A similar small-angle departure from the Hellings-Downs curve occurs due to the pulsar terms~\cite{Mingarelli:2014xfa}.  Note that equality between EE and BB can not be used as a null test of general relativity at small angular separations.
\begin{figure}[htp]
\includegraphics[clip=true,angle=0,width=0.5\textwidth]{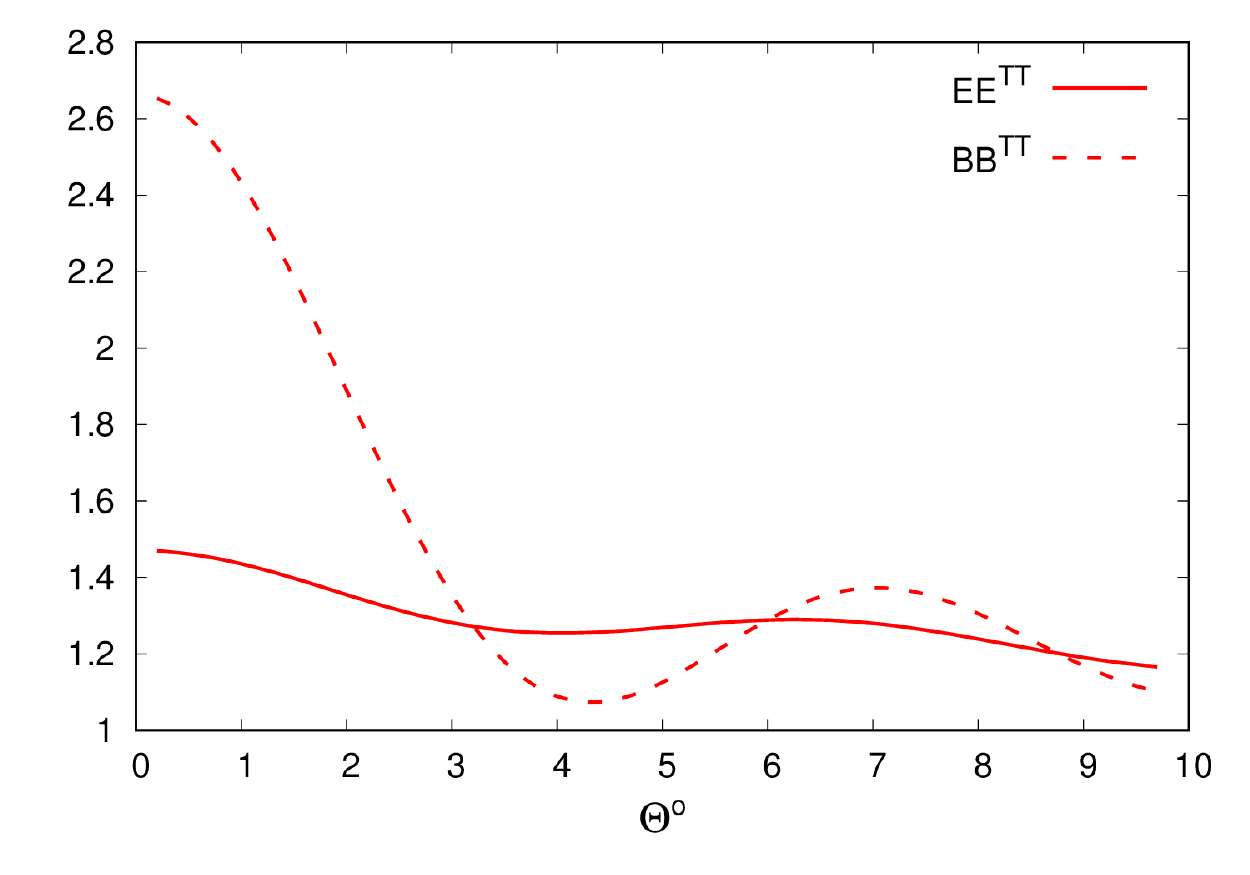} 
 \vspace*{-0.2in}
\caption{\label{fig:small} The EE and BB correlation functions for the TT mode at small angular separations. The star terms break the equality between EE and BB. The plots are shown for $fL=10$.}
\end{figure}
Second, we find that the VL correlations are enhanced by a leading factor that is logarithmic  in $fL$:
\begin{eqnarray}
&& {\rm EE}^{\rm VL}(\Phi_{s})|_{\Theta=0} =  2\,\text{ln}(2\,\Phi_{s}) -\frac{38}{15} + 2\gamma_{E} + \mathcal{O}\Big(\frac{1}{\Phi_{s}} \Big)  \nonumber\\
&& {\rm BB}^{\rm VL}(\Phi_{s})|_{\Theta=0} =   2\,\text{ln}(2\,\Phi_{s}) -\frac{14}{3} + 2\gamma_{E}  +  \mathcal{O}\Big(\frac{1}{\Phi_{s}} \Big)   \, .
\end{eqnarray}
Third, the SL modes are enhanced by a leading factor that is linear in $fL$:
\begin{eqnarray}
	&& {\rm EE}^{\rm SL}(\Phi_{s})|_{\Theta=0} = 
	\frac{1}{2} \Big[ -\frac{19}{10} + 3 \gamma_{E} +\frac{\pi}{3}\Phi_{s}\nonumber \\
	&& \hspace*{1.3in} - \text{log}(2\,\Phi_{s}) + \mathcal{O}\Big(\frac{1}{\Phi_{s}} \Big)  \Big],
\end{eqnarray}
where $\gamma_{E} $ is the Euler-Gamma constant. The enhanced response to longitudinal gravitational waves will allow us to place far more stringent bounds on the energy density in these modes. The majority of stars in the Gaia catalog are at distances of $L=0.1 \rightarrow 10$ kpc~\cite{0004-637X-833-1-119} and the duration and cadence of the Gaia mission will allow us to probe gravitational frequencies between $3\times 10^{-9}$ Hz and $5\times 10^{-7}$ Hz~\cite{PhysRevLett.119.261102}. Thus the typical values for $fL$ will range between 30 and $5\times 10^5$, and the bounds for the SL mode in particular should be orders of magnitude tighter than for the usual TT modes of GR.

\section{Discussion}

Astrometry is a promising new approach for detecting gravitational waves~\cite{Book:2010pf, Klioner:2017asb, PhysRevLett.119.261102}. Astrometry is particularly well suited to constraining the polarization content of gravitational waves. The expressions for the two-point correlation functions and the E and B mode correlation patterns that we have computed here provide a starting point for the analysis of data from the Gaia mission~\cite{2016A&A...595A...1G}. Implementing such an analysis will be challenging: our expressions for the EE and BB correlations assumed a uniform distribution of stars on the sky, and we placed all the stars at the same distance from the Earth. In reality the Gaia's sky coverage will be non-uniform in location and distance, and techniques similar to those used in cosmic microwave background analyses will have to be used to extract the uniform correlation patterns from experiments with non-uniform sky coverage~\cite{Hinshaw:2003ex, Mignard:2012xm}. The vanishing of the BB correlations for scalar polarizations and the near-equality of the EE and BB polarization modes for general relativity provides a potentially powerful null test. It will be interesting to assess how the power of this test is impacted by non-uniform sky coverage and observation noise. In addition to measuring the locations of billions of stars in our galaxy, Gaia will also measure the locations of thousands of quasars. These quasars are at vastly larger distances, making the $fL$ dependent enhancements for the longitudinal modes much more significant. Pairs of nearby quasars, or individual strongly lensed quasars that produce multiple images, may well provide the strongest limits to the energy density of longitudinal polarization states.

\section*{Acknowledgments}
We appreciate the support of the NSF Physics Frontiers Center Award PFC-1430284.

\bibliography{refs}

\end{document}